# The energy limit in molecular alignment by femtosecond laser pulse

Yue-Xun Li, Tao Du

Department of Physics, Yunnan Minzu University, Kunming 650500, People's Republic of China

yuexunli@ymu.edu.cn

Abstract:

In this paper, we studied the optimal alignment and anti-alignment of $O_2$ molecules by femtosecond laser pulse under different input energy. The results show that there is an energy limit in molecular alignment. The optimal molecular alignment and anti-alignment degree will never increase with the increase of laser energy when the input laser energy exceeds the energy limit. Besides, under the energy limit, the molecular optimal alignment (or anti-alignment) degree by using a shorter pulse duration (under 100 fs) will be about same when the input laser energy remained constant (different pulse duration and the peak power are chosen). Especially at a lower temperature of molecular assemble.

1. Introduction

Laser-induced molecular alignment and orientation play a significant role in ultra-fast strong laser technology [1][2][3][3][5], chemistry and chemical reaction dynamics [6][7][8][9] and so on. There is great development in femtosecond induced free-field molecular alignment in nearly 30 years. And these developments focus on enhancement and control of molecular alignment and orientation by using all kinds of laser technologies[10][11][12][13], the population control of rotational states [14][15][16][17][18][19][20], and the further application of aligned and oriented molecules, such as the generation of high harmonic[1][2][3] and attosecond laser pulse from aligned molecules[4][5], the tomography of electronic or molecular orbits by using aligned molecules[3][21][22][23], chemical reaction control [9][24] and molecular nanometer design [25][26][27].

Molecular alignment and orientation can be achieved in both adiabatic and nonadiabatic regimes. Due to its advantage, nonadiabatic alignment has attracted the interests of both physicists and chemists. Molecular alignment can be obtained under free-field condition for nonadiabatic alignment. In many documents, it is found that the molecular alignment will be enhanced with the increase of the peak power and duration of laser pulse if the molecules haven't been ionized. However, there are no studies which discuss the exact peak power and duration of laser pulse we should use in aligning molecules.

In this paper, we illustrate the $O_2$ molecules alignment by femtosecond laser with different input laser pulse energy (different peak power and pulse duration are chosen) and temperature. From the calculated results we found that there is an energy limit for $O_2$ alignment by using femtosecond no matter the temperature of molecular ensemble. Besides, when we remain the input energy constant by choosing different pulse duration and the peak power, the optimal alignment (or anti-alignment) degree will be about same. Especially at a lower temperature of molecular assemble and the pulse duration is shorter (under 100fs).

2. Theoretical model

We calculate the non-adiabatic alignment of $O_2$ molecules assemble using the method presented in Ref.[28][29]. Briefly, non-adiabatic field-free alignment is achieved by the excitation of a rotational wave packet, which is represented by the spherical harmonics under the rigid rotor approximation

$$\psi(t) = \sum_{J,M} A_{J,M}(t) \left| J,M \right\rangle \tag{1}$$

The time evolution of the wave packet can be calculated through solving the time-dependent Schrödinger equation (TDSE)

$$i\frac{\partial \psi(\theta,t)}{\partial t} = [B\boldsymbol{J}^2 + H_{\text{int}}]\psi(\theta,t) \tag{2}$$

in which the θ is defined by the angle between the laser field polarization and molecular axis, and the $B\boldsymbol{J}^2$ is the rotational energy operator. The Hamiltonian of the angle-dependent AC stark shift

$$H_{\text{int}}(\theta, t) = -\frac{1}{4}\Delta\alpha\varepsilon_0^2 f^2(t)\cos^2\theta \tag{3}$$

Where $\Delta\alpha = \alpha_{\|} - \alpha_{\perp}$, $\alpha_{//}$ and $\alpha_{\perp}$ are the longitudinal and transverse components of the polarizability tensor, $\theta$ is the angle between the molecular axis and the direction of the laser polarization, while $\varepsilon_0$ and $f(t)$ are the electric field magnitude and the envelope function of the linearly-polarized laser pulse. Finally, the time dependent alignment parameter is obtained by

$$< \cos^2\theta(t) > = \left\langle \Psi(t) \middle| \cos^2\theta \middle| \Psi(t) \right\rangle \tag{4}$$

In our calculation, the $O_2$ ($B_0$ = 1.4297cm$^{-1}$, De = 4.839 × 10$^{-6}$cm$^{-1}$, and Δα = 1.099Å$^3$[30]) is aligned by femtosecond (fs) laser pulses (choose different peak power and pulse duration) with a Gaussian temporal.

3. Results and analysis

(1) The energy limit in molecular alignment

We calculate the molecular alignments when the laser pulse durations are 5fs and from 25fs to 200fs (increase by 25fs), the peak powers of laser pulse are from 10TW to 150TW (increase by 20TW), and the temperatures of molecular assemble are 10K, 80K and 300K respectively. We only choose those calculated results which belong to non-adiabatic molecular alignment. The results are shown in the fig. 1.

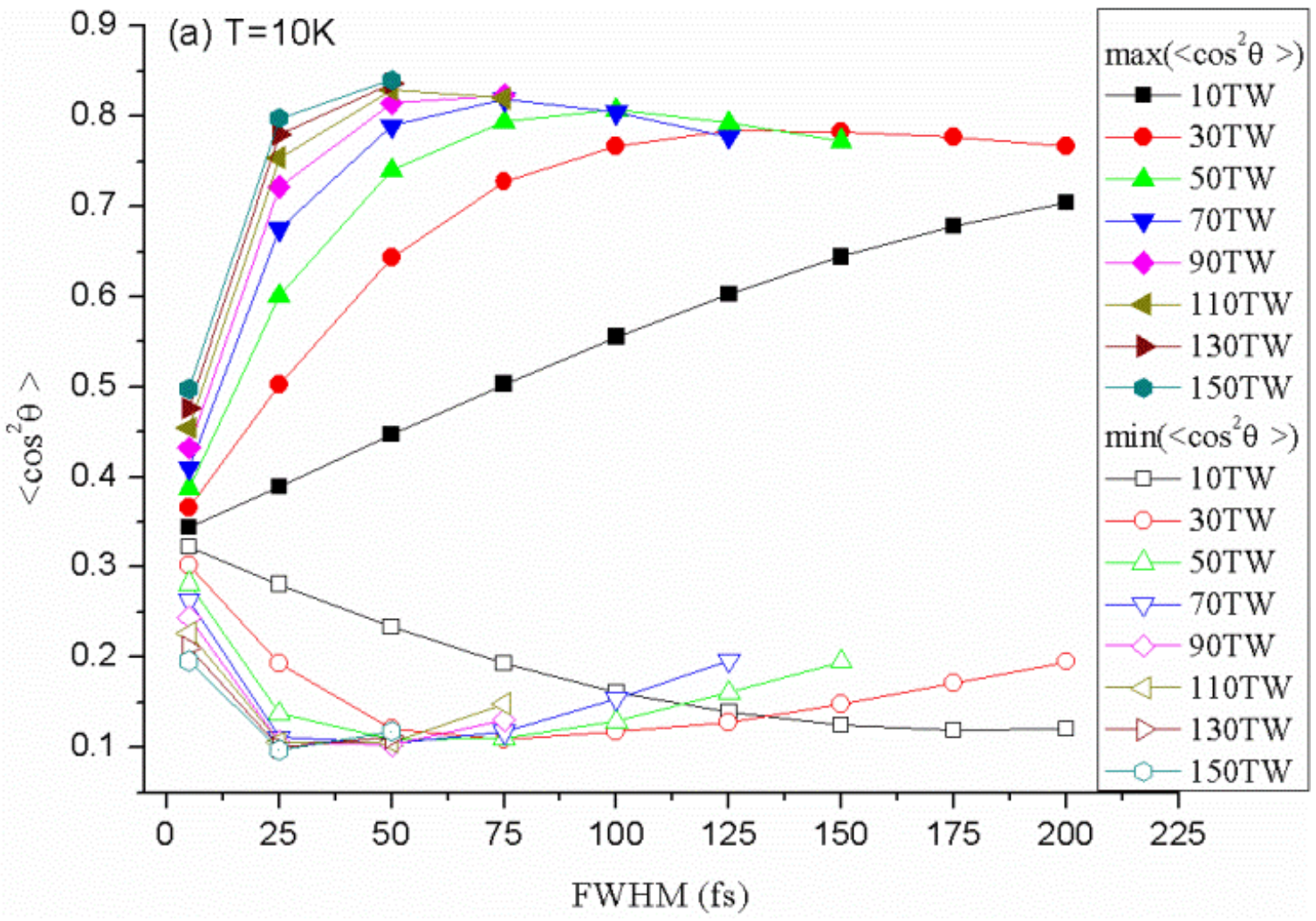

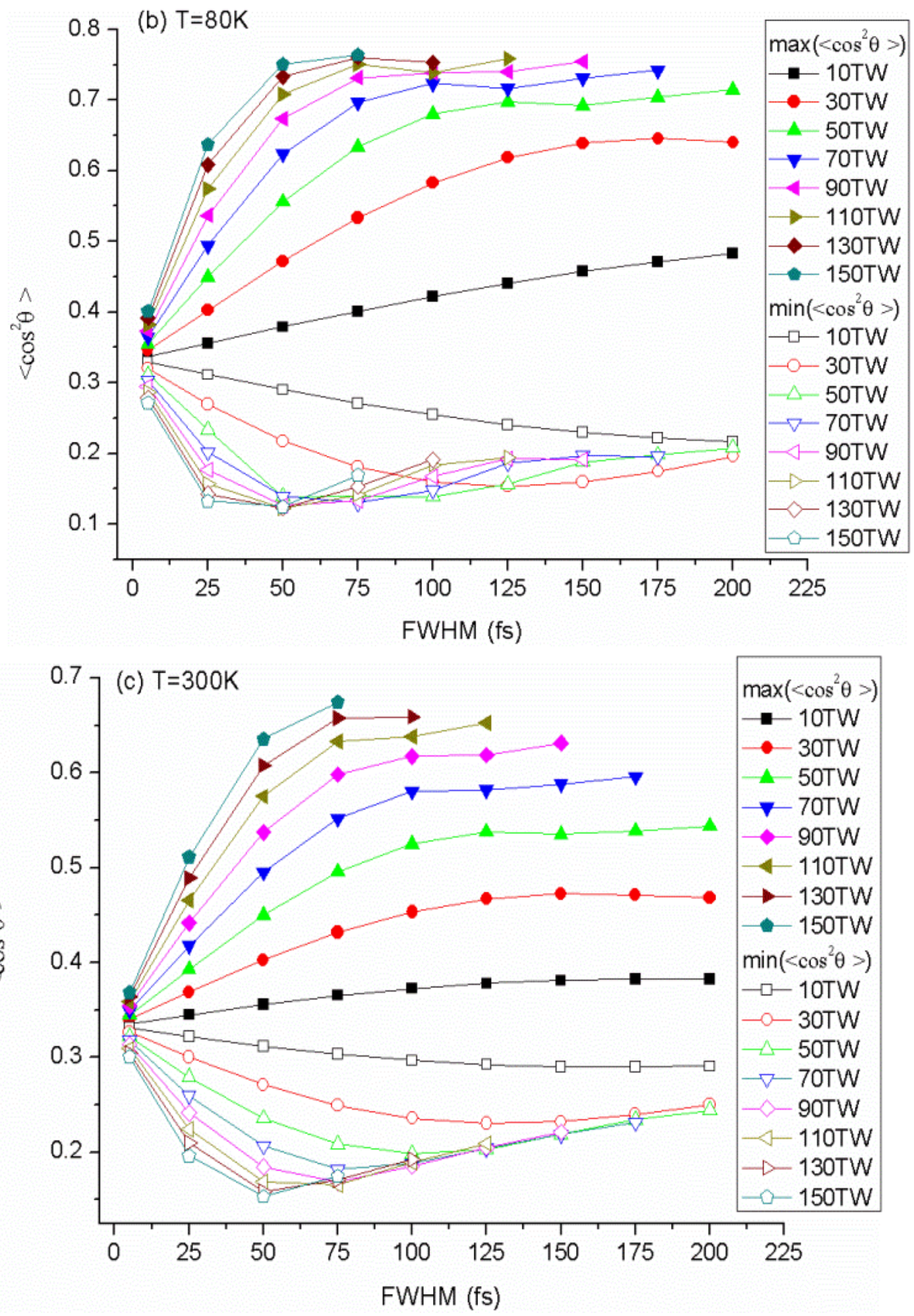

Fig. 1. The optional alignment (max( $< \cos^2\theta >$ )) and anti-alignment (min( $< \cos^2\theta >$ )) of $O_2$ molecules with different input energies when the temperature are: (a)T=10K; (b)T=80K; (C)T=300K.

From the fig. 1, we can see that the optimal molecular alignment and anti-alignment will increase with the increase of input energy when the input energy is not high. But it will remain unchanged and even decrease when the input energy exceeds to certain energy (we called it energy limit in this paper). The approximate energy limits are as follows: 6.5J for alignment and 2.5J for anti-alignment under 10K in fig. 1(a), 6.5 J for alignment and 3.0J for anti-alignment under 80K in 1(b), and 6.5J for both alignment and anti-alignment under 300K in fig. 1(c). For example, in fig. 1(a), when the pulse duration is 50fs, the maximum peak power of laser pulse is around 130TW for optimal alignment and 50TW for optimal anti-alignment; but when the pulse duration is 100fs, the maximum peak power is around 50TW (or more but less than 70TW) for optimal alignment and 30 TW for optimal anti-alignment. In fig. 1(C), the maximum peak power is around 130TW for both the optimal alignment and anti-alignment when the pulse duration is 50fs.

We can also see that the optimal alignment and anti-alignment in fig.1 (b) and (c) will change more slowly than the one in fig. 1(a) when the pulse durations are 5fs and 25fs. The reason is that the molecular rotational states are basically populated in lower energy levels when the temperature are lower (T=10K in fig. 1(a)). And the energy level intervals are smaller at the lower energy levels than the ones at the higher levels. Therefore, molecular rotational states can be transited easily when they populated in lower energy levels. However, when temperature is higher (in fig. 1(b) and (c)), the molecular energy levels are populated much wider energy range. Some rational states are even populated in higher level and the according energy intervals are much bigger. The input laser pulse with a shorter duration makes the molecular transit difficultly because it doesn't last enough long. Besides, under a higher temperature (in fig.1(c), T=300K), the optimal alignment and anti-alignment will be increase regularly with the increase of peak power of input laser pulse under energy limit. But under a lower temperature (in fig.1 (a), T=10K), the optimal alignment and anti-alignment will become increase slowly with the increase of peak power of input laser pulse.

(2) Some rules under the energy limit

Form the above calculated results, we can conclude that there is an energy limit in molecular alignment. When the input energy is over the energy limit, the optimal molecular alignment and anti-alignment cannot increase and even decrease with the increase of input energy. So we should choose an appropriate input energy which is under the energy limit when we align molecules by using femtosecond. As far as O2 molecules are concerned, the energy limit is around 6.5J. Therefore, the input energy we choose for its aligning is under 6.5J. The exact pulse duration and peak power depend on the temperature. For example, the maximum peak power should be 130TW if we choose 50 fs when temperature is 80K.

In the following calculation, we calculate the optimal alignment and anti-alignment at an input energy under and little over the energy limit when the temperature is 10K, 80K and 300K. The results are shown in the fig. 2.

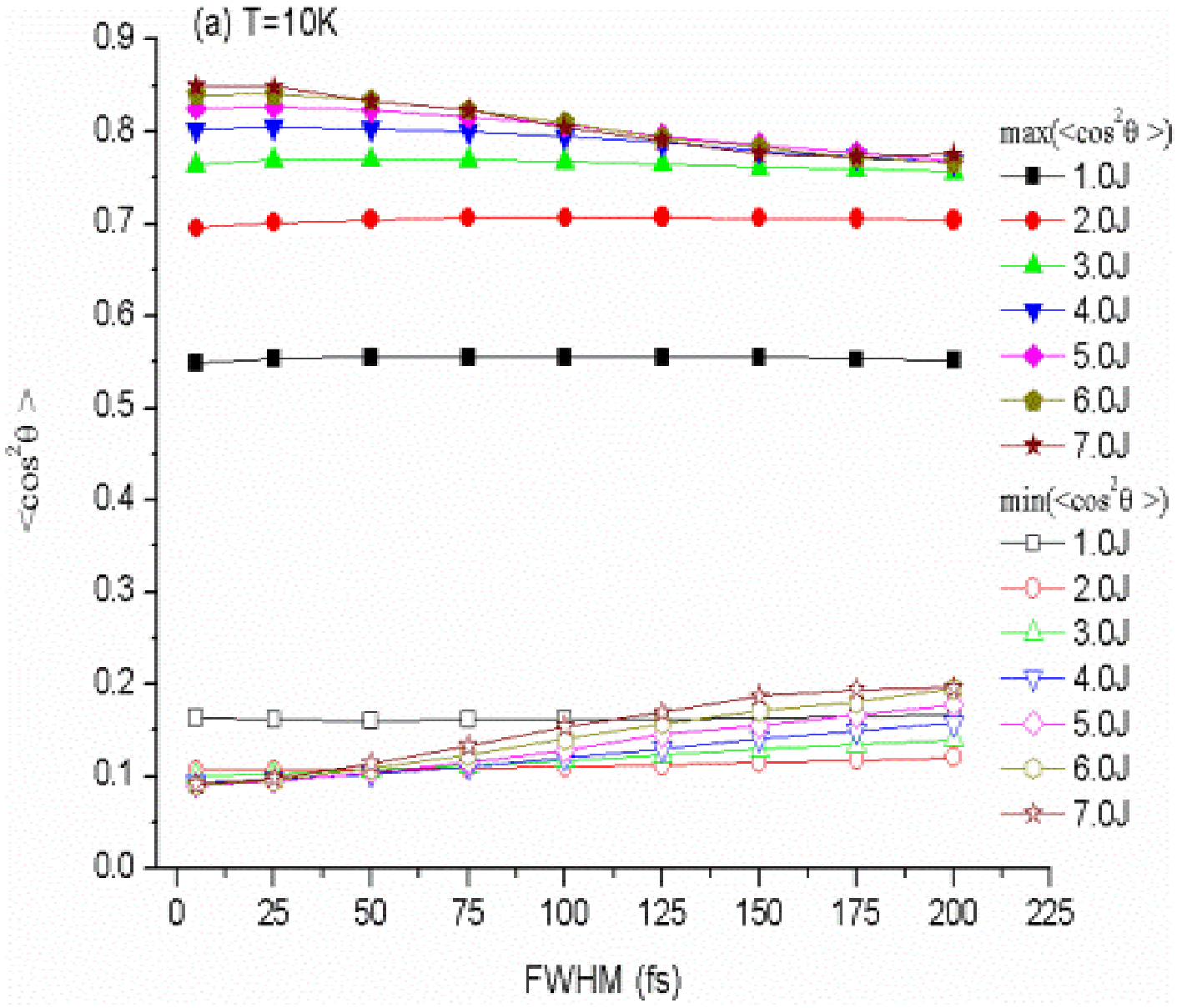

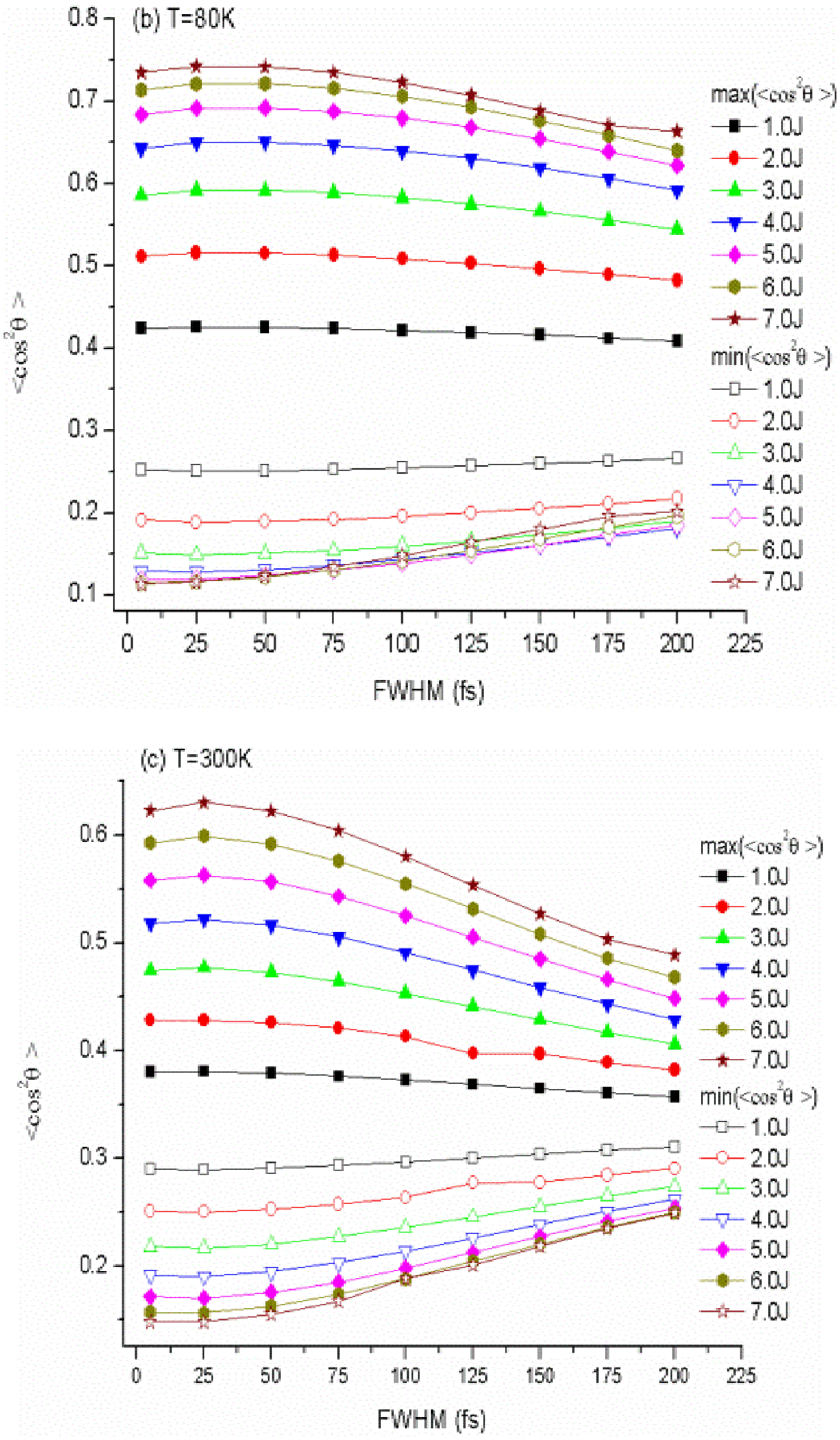

Fig. 2. The optional alignment (max($<\cos^2\theta>$)) and anti-alignment (min($<\cos^2\theta>$)) of $O_2$ molecules at the same input energies (under and little over the energy limit). (a)T=10K; (b)T=80K; (c)T=300K.

It can be shown in fig. 2 that under the energy limit the optional molecular alignment and anti-alignment by femtosecond laser pulses with same input energy (deferent pulse duration and peak power) are nearly same. Especially when the temperature is lower (in fig.2 (a) and (b)) or the input energy is lower (<4J). Besides, there are some differences when the temperatures are different. Firstly, the optimal alignment and anti-alignment under the same input energy will be gradually decrease with the increase of pulse duration (the peak power will be decrease since keeping the same input energy). And this trend is more obvious when the temperature is high (shown in fig. 2(b) and (c)). Secondly, the increment of alignment and anti-alignment degree will dramatic reduce with increasing of input energy (under energy limit).

From the above results, it can be seen that there is energy limit existed in molecular alignment by using femtosecond laser pulse. Besides, under the energy limit, the optional molecular alignment and anti-alignment will be around same as the same input energy when the temperature is lower or else the laser pulse duration is in a shorter duration range. Molecular alignment and anti-alignment will decrease with the increase of laser pulse duration especially at higher temperature. The reasons can be explained as follows: the laser pulse gives molecules a kick when a femtosecond aligns molecules. The pulse duration is the lasting time of the kick and the peak power is the strength of the kick. Therefore, molecular alignment will be increase no matter increase the peak power or pulse duration of laser pulse. But the strength of the kick plays a signature role in molecular alignment[16][31], which gives the molecules enough force and makes the molecules transmit from the lower levels to the higher levels. To keep the same input energy, the peak power will be low when the pulse duration is long. Therefore, the all rotational states (especially the higher rotational states) cannot acquire an enough kick to transmit another level when the peak power is low at long pulse duration. So the optimal alignment and anti-alignment will be decrease when pulse duration increase (the peak power will decrease for keeping a same input energy).

In order to verify whether the conclusions under the energy limit is still right or not in two laser pulses alignment, we also calculate the two laser pulses alignment. The parameters are as follows: T=80K, $E_{total}$=4.0J, the ratio and delay of two pulses are chosen according to slope rule[13]. So the parameters of pulse are as follows: $E_1/E_2$=1.0 and the second pulse was inserted at the maximum of slopes of the alignment signal by the first laser pulse. The results calculated by one laser pulse (E=4.0J) and two pulses ($E_{total}$=4.0J, $E_1/E_2$=1.0) are shown in fig. 3.

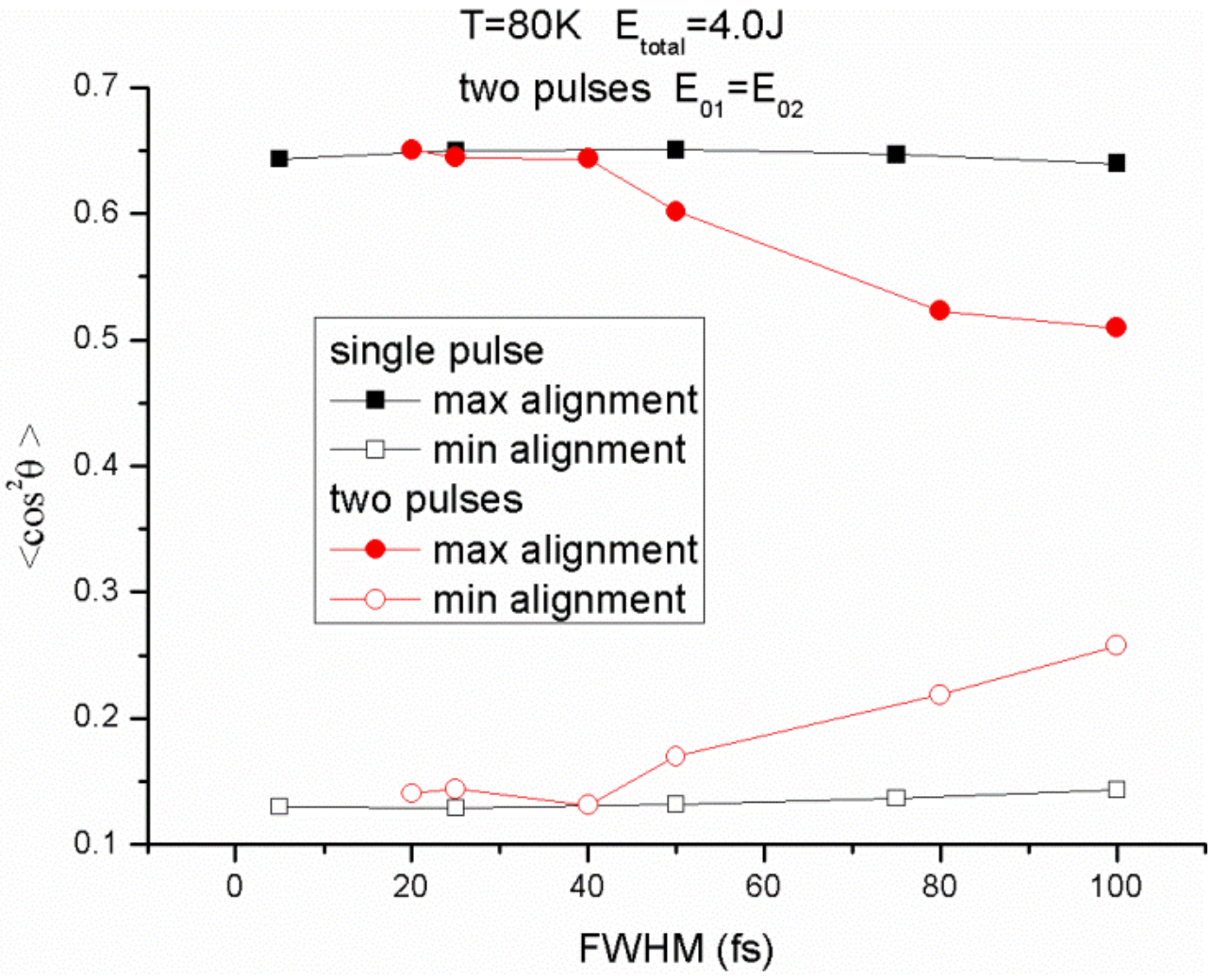

Fig. 3. The optional alignment and anti-alignment of O2 molecules under one pulse (black line) and two pulses (red line) at the same input energy.

From fig. 3, we can see that the optimal alignment and anti-alignment will be around same when the pulse duration is under 50fs and decrease when the pulse duration is longer than 50fs. And it will be gradually decrease with the increase of pulse duration. So molecular alignment and anti-alignment will be around same when the duration of laser pulse is short (under 50fs) no matter molecular alignment by one pulse or two pulses. And the optional alignment and anti-alignment will be decrease with the increase of pulse duration. Therefore, we should choose a propitiate pulse duration and peak power under energy limit in molecular alignment.

## 4. Conclusion

This paper studied theoretically the optimal alignment of $O_2$ molecules with the increase of femtosecond laser energy. The results show that there are an energy limit existed which is the maximum input energy for molecular alignment. It is useless to molecular alignment when the input laser pulse energy exceed to the energy limit. Besides, under the energy limit molecules will acquire an around same optimal alignment and anti-alignment degree with same input energy (although they have different pulse duration and peak power) when the pulse duration is under 100 fs. The molecular alignment and anti-alignment degree will be decrease with the increase of pulse duration (the peak power will be increase at the same input energy). Finally, the molecular alignment by two laser pulses which have a same input energy as one laser pulse will also have an around same optimal alignment and anti-alignment degree as the one by one laser pulse. However, which is satisfied just under 50 laser pulse duration.

Acknowledge

This work was supported by the National Natural Science Foundation of China (Grant Nos.61640415 and 11964042) and China Scholarship Council.

References;